\documentclass[showpacs,aps,twocolumn]{revtex4}
\usepackage{epsfig}
\usepackage{graphicx}
\usepackage{amsmath,amssymb,amsfonts}
\usepackage{array}
\usepackage{url}
\usepackage{multirow}
\usepackage{natbib}
\usepackage{float}
\usepackage{lineno}
\usepackage{xspace}
\usepackage{ulem}

\newcommand{\bef}{\begin{figure}}
\newcommand{\eef}{\end{figure}}
\newcommand{\bc}{\begin{center}}
\newcommand{\ec}{\end{center}}

\newcommand{\be}{\begin{equation}}
\newcommand{\ee}{\end{equation}}
\newcommand{\bea}{\begin{eqnarray}}
\newcommand{\eea}{\end{eqnarray}}

\def\ba{\begin{eqnarray}}
\def\ea{\end{eqnarray}}
\usepackage[usenames,dvipsnames]{color}
\definecolor{darkblue}{RGB}{0,0,196}
\usepackage[colorlinks=true,linkcolor=darkblue,citecolor=darkblue,urlcolor=darkblue]{hyperref}

\begin{document}
\title{Conductivity, diffusivity, and violation of Wiedemann-Franz Law in a hadron resonance gas with van der Waals interactions}
 
\author{Kshitish Kumar Pradhan}
\author{Dushmanta Sahu}
\author{Ronald Scaria}
\author{Raghunath Sahoo\footnote{Corresponding Author Email: Raghunath.Sahoo@cern.ch}}
\affiliation{Department of Physics, Indian Institute of Technology Indore, Simrol, Indore 453552, India}

\begin{abstract}

In this work, a hadron resonance gas under van der Waals (VDW) interactions has been studied. Both attractive and repulsive interactions between the meson-meson and (anti)baryon-(anti)baryon have been taken into consideration. Various transport properties such as electrical conductivity ($\sigma_{\rm el}$) and thermal conductivity ($\kappa_{\rm th}$) have been estimated by solving the Boltzmann transport equation under the relaxation time approximation. The effect of baryochemical potential ($\mu_{\rm B}$) and temperature is also explicitly explored for the mentioned observables.  Comparisons have been made with the results obtained from other existing theoretical models. We observe the violation of Wiedemann-Franz law in a hadron resonance gas at a high-temperature regime. The corresponding diffusivities have also been estimated, which can help us to understand the systems better.

\end{abstract}
\date{\today}
\maketitle

\section{Introduction}
\label{intro}

Various transport properties can act as necessary probes to analyze the systems formed in ultra-relativistic collisions. Transport coefficients such as shear viscosity ($\eta$), bulk viscosity ($\zeta$) along with the thermal conductivity ($\kappa_{\rm th}$) and electrical conductivity ($\sigma_{\rm el}$) contribute to the dissipative nature of any fluid. These coefficients are expressed in terms of the distribution function's responses to the non-uniformities present in the system. More importantly, in the hot and dense system formed in heavy-ion collisions where the strong interaction is expected to be prevalent, these coefficients can have a substantial role in determining the behavior of system evolution. They can also help us locate the phase transition through their dependencies on various system parameters such as temperature and chemical potential. 

In peripheral heavy-ion collisions, a large transient electromagnetic field is expected to be formed \cite{Voronyuk:2011jd, Tuchin:2013ie}. The strong magnetic field decays with the space-time evolution of the system. However, the rapid decay in the magnetic field is slowed down due to the induced electric current. This can be quantified by the electrical conductivity ($\sigma_{\rm el}$), which measures the system's response to the small perturbations from the electromagnetic field. $\sigma_{\rm el}$ can have a considerable effect in the enhancement of the low mass dilepton production as well as in the soft photon production \cite{Akamatsu:2011mw, Yin:2013kya}. It is also related to the charge diffusion coefficient through the well-known relation $D_{el} = \sigma_{el}/\chi_{Q}$, where $\chi_{Q}$ is the charge susceptibility \cite{Aarts:2014nba}. The electrical conductivity is also partially responsible for the chiral magnetic effect, which is a signature of $\mathcal C \mathcal P$ violation \cite{Fukushima:2008xe}. Although direct experimental measurement of $\sigma_{\rm el}$ is not possible, a myriad of theoretical and phenomenological studies have been conducted to estimate this coefficient. It has been estimated from the Monte-Carlo simulations in lattice QCD (LQCD) calculations \cite{Aarts:2020dda}. In Ref. \cite{Cassing:2013iz}, the scaled electrical conductivity estimated from the parton-hadron-string dynamic (PHSD) approach shows a minimum near the critical temperature. In addition, $\sigma_{\rm el}$ has been estimated from the kinetic theory approach \cite{MartinezTorres:2014son}, holography \cite{Finazzo:2013efa}, Dyson-Schwinger calculation \cite{Qin:2013aaa}, excluded volume hadron resonance gas model \cite{Kadam:2017iaz}, color string percolation model \cite{Sahoo:2018dxn}, and quasi-particle model \cite{Srivastava:2015via}. 

Along with this, another transport coefficient that plays a crucial role in the thermodynamic evolution of the system is the thermal conductivity ($\kappa_{\rm th}$). The Fourier law of thermal conduction connects the heat flux to the temperature gradient and the thermal conductivity. It is also very closely related to the isospin diffusivity in the nuclear matter \cite{Shi:2003np}. In high baryon-rich environments, such as the systems produced in FAIR and NICA, $\kappa_{\rm th}$ has of significant importance in the hydrodynamic evolution. It also diverges near the critical temperature and can be used to study the hydrodynamic fluctuations on experimental observables by the fluctuation-dissipation theorem \cite{Kapusta:2012zb}. This coefficient has also been studied extensively by semiclassical transport theory \cite{Mitra:2017sjo}, excluded volume hadron resonance gas model \cite{Kadam:2017iaz}, color string percolation model \cite{Sahoo:2019xjq} and Nambu-Jona-Lasinio model \cite{Marty:2013ita}.

The lattice QCD (lQCD) approach has been very effective in estimating the thermodynamic properties of the strongly interacting matter \cite{Borsanyi:2013bia, HotQCD:2014kol}. In such studies, a rapid crossover phase transition can be observed between the temperature range, $T \sim$ 140-190 MeV \cite{Aoki:2006we}. While the lQCD calculations give effective results at zero or low baryochemical potential, interacting hadron gas models can explore baryon-rich environments where $\mu_{\rm B}$ is very high. Such treatment will give us exciting insights into the systems that are formed in the hadronic phase of the evolution at the LHC, RHIC, FAIR, and NICA. Although at a low-temperature regime ($T \sim$ 100-150), the Hadron Resonance Gas (HRG) model can explain the lattice QCD data very well, the HRG model predictions at high temperatures deviate from that of the lattice QCD. This can be attributed to the statement that hadrons quickly melt after temperature $T \sim$ 160 MeV. However, recent works have made some substantial efforts to show that this conclusion is very premature because the van der Waals interaction can play an essential role in determining the thermodynamics of the hadron gas at high temperatures \cite{Vovchenko:2016rkn}. In the ideal HRG model, where there is no interaction involved, estimations of the higher-order charge fluctuations substantially deviate from that of the lQCD estimations \cite{Pal:2020ucy}. Once we include interactions into the ideal HRG model, the model becomes sensitive to the higher-order charge fluctuations, and at zero baryochemical potential, it gives results that match with lQCD results to a reasonable extent.

This work studies a hadron gas subjected to van der Waals (VDW) interactions, both attractive and repulsive, and estimates various transport coefficients by solving the Boltzmann transport equation using a relaxation time approximation. We take the (anti)baryon-(anti)baryon and meson-meson interactions into account, while the meson-(anti)baryon and baryon-antibaryon interactions are neglected for the sake of simplicity. We present a few results showing the behavior of conductivities and diffusivities under the change in both temperature and baryon density. The paper is organized as follows. In section \ref{formulation}, a detailed formulation of electrical and thermal conductivity in relaxation time approximation for a van der Waals hadron resonance gas is given. Section \ref{results} discusses the obtained results, and finally, all the results are summarized in section \ref{sum}.

\section{Formulation}
\label{formulation}

\subsection{van der Waals HRG model}
\label{ss1}

Under the ideal HRG model, the hadrons are considered non-interacting point particles. However, nucleon-nucleon scattering experiments suggest the existence of finite van der Waals (VDW) interactions. Taking this into consideration, the van der Waals HRG (VDWHRG) model was developed to take both attractive and repulsive hadronic interactions. 

The partition function for $i^{th}$ particle species in a Grand Canonical Ensemble (GCE) of ideal HRG can be written as~~\cite{Andronic2012}
\begin{equation}
\label{eq1}
ln Z^{id}_i = \pm \frac{Vg_i}{2\pi^2} \int_{0}^{\infty} p^2 dp\ ln\{1\pm \exp[-(E_i-\mu_i)/T]\}.
\end{equation}
Here, $g_i$ and $E_i = \sqrt{p^2 + m_i^2}$ are defined as the degeneracy and energy of the $i^{th}$ hadron, respectively. The $\pm$ sign correspond to fermions and bosons respectively.  $\mu_{i}$ denotes the corresponding chemical potential, which is given by 
\begin{equation}
\label{eq2}
\mu_i = B_i\mu_B + S_i\mu_S +Q_i\mu_Q,
\end{equation}
where baryon chemical potential, strangeness chemical potential and charge chemical potential are given by $\mu_{B}$, $\mu_{S}$ and $\mu_{Q}$, respectively. $B_i$, $S_i$ and $Q_i$ denote the baryon number, strangeness and electric charge of the $i^{th}$ hadron. Pressure $P_i$, energy density $\varepsilon_i$, number density $n_i$ and entropy density $s_i$ can now be obtained from the partition function as 
\begin{equation}
\label{eq3}
P^{id}_i(T,\mu_i) = \pm \frac{Tg_i}{2\pi^2} \int_{0}^{\infty} p^2 dp\ ln\{1\pm \exp[-(E_i-\mu_i)/T]\}
\end{equation}
\begin{equation}
\label{eq4}
\varepsilon^{id}_i(T,\mu_i) = \frac{g_i}{2\pi^2} \int_{0}^{\infty} \frac{E_i\ p^2 dp}{\exp[(E_i-\mu_i)/T]\pm1}
\end{equation}
\begin{equation}
\label{eq5}
n^{id}_i(T,\mu_i) = \frac{g_i}{2\pi^2} \int_{0}^{\infty} \frac{p^2 dp}{\exp[(E_i-\mu_i)/T]\pm1}
\end{equation}
\begin{align}
 s^{id}_i(T,\mu_i)=&\pm\frac{g_i}{2\pi^2} \int_{0}^{\infty} p^2 dp \Big[\ln\{1\pm  \exp[-(E_i-\mu_i)/T]\}\nonumber\\ 
&\pm \frac{(E_i-\mu_i)/T}{\exp[(E_i-\mu_i)/T]\pm 1}\Big]
 \label{eq6}
 \end{align}
 In a canonical ensemble representation, the VDW equation can be written as \cite{Samanta:2017yhh,W.Greiner123Stocker}
 \begin{equation}
\label{eq7}
    \Bigg( P + \bigg(\frac{N}{V}\bigg)^{2}a\Bigg)\big(V-Nb\big) =  N T
\end{equation}
where a and b (both positive) are the usual VDW parameters describing the attractive and repulsive interactions, respectively. Pressure, volume, temperature and number of particles in the system are denoted by P, V, T and N, respectively.

In terms of number density, $n \equiv N/V,$ the above equation can be simplified as
\begin{equation}
\label{eq8}
    P(T,n) = \frac{nT}{1-bn}- an^{2},
\end{equation}
The correction due to repulsive interactions is included in the first term of Eq.\ref{eq8} by replacing the total volume V with the effective volume available to particles using the proper volume parameter $b = 16\pi r^{3}/3$, $r$ being the particle hardcore radius. The second term takes care of attractive interactions between particles.

The VDW equation of state in GCE can then be written as \cite{Samanta:2017yhh,Vovchenko:2015vxa,Vovchenko:2015pya} 
\begin{equation}
\label{eq9}
    P(T,\mu) = P^{id}(T,\mu^{*}) - an^{2}(T,\mu),
\end{equation}
where the n(T,$\mu$) is the particle number density of the VDW hadron gas and is given by
\begin{equation}
\label{eq10}
    n(T,\mu) = \frac{\sum_{i}n_{i}^{id}(T,\mu^{*})}{1+b\sum_{i}n_{i}^{id}(T,\mu^{*})}.
\end{equation}
Here, $i$ runs over all hadrons in the interacting medium and $\mu^{*}$ is the modified chemical potential given by 
\begin{equation}
\label{eq11}
    \mu^{*} = \mu - bP(T,\mu) - abn^{2}(T,\mu) + 2an(T,\mu).
\end{equation}
Other thermodynamical variables like entropy density $s(T,\mu)$ and energy density $\epsilon(T,\mu)$ can now be obtained as,
\begin{equation}
\label{eq12}
s(T,\mu) = \frac{s^{id}(T,\mu^{*})}{1+bn^{id}(T,\mu^{*})},
\end{equation}
\begin{equation}
\label{eq13}
\epsilon(T,\mu) = \frac{\sum_{i}\epsilon_{i}^{id}(T,\mu^{*})}{1+b\sum_{i}n_{i}^{id}(T,\mu^{*})} - an^{2}(T,\mu).
\end{equation}

The VDWHRG model, in its initial formalism, includes interactions only between all pairs of baryons or all pairs of anti-baryons \cite{Samanta:2017yhh,Vovchenko:2015vxa,Vovchenko:2015pya,Vovchenko:2016rkn}. The interaction between baryon-antibaryon pairs was neglected as the short-range interactions between them are believed to be dominated by annihilation processes \cite{Andronic2012,Vovchenko:2016rkn}. Meson-meson or meson-(anti)baryon interactions are neglected as the inclusion of mesonic interactions leads to a suppression of thermodynamical quantities in the crossover region at $\mu_{B} = 0.0$ GeV in comparison with LQCD data \cite{Vovchenko:2016rkn}. However, a formalism including meson-meson interactions was developed by choosing the VDW parameters which best fit the LQCD data~\cite{Sarkar:2018mbk} by considering a hardcore radius $r_{M}$ for mesons.      
Hence the total pressure in the VDWHRG model is modified as \cite{Samanta:2017yhh,Vovchenko:2015vxa,Vovchenko:2015pya,Vovchenko:2016rkn,Sarkar:2018mbk} 
\begin{equation}
\label{eq14}
P(T,\mu) = P_{M}(T,\mu) + P_{B}(T,\mu) + P_{\bar{B}}(T,\mu),
\end{equation}
where the $P_{M}(T,\mu), P_{B(\bar B)}(T,\mu)$ are the contributions to pressure from mesons and (anti)baryons, respectively and are given by,
\begin{equation}
\label{eq15}
P_{M}(T,\mu) = \sum_{i\in M}P_{i}^{id}(T,\mu^{*M}),       
\end{equation}
\begin{equation}
\label{eq16}
P_{B}(T,\mu) = \sum_{i\in B}P_{i}^{id}(T,\mu^{*B})-an^{2}_{B}(T,\mu),
\end{equation}
\begin{equation}
\label{eq17}
P_{\bar{B}}(T,\mu) = \sum_{i\in \bar{B}}P_{i}^{id}(T,\mu^{*\bar{B}})-an^{2}_{\bar{B}}(T,\mu).
\end{equation}
Here, $M$, $B$ and $\bar B$ represent mesons, baryons and anti-baryons, respectively. $\mu^{*M}$ is the modified chemical potential of mesons because of the excluded volume correction and $\mu^{*B}$ and $\mu^{*\bar B}$ are the modified chemical potentials of baryons and anti-baryons due to VDW interactions~\cite{Sarkar:2018mbk}. Considering the simple case of vanishing electric charge and strangeness chemical potentials \cite{Bazavov:2017dus}, $\mu_{Q} = \mu_{S} = 0$, the modified chemical potential for mesons and (anti)baryons can be obtained from Eq.~\ref{eq2} and Eq.~\ref{eq11} as; 
\begin{equation}
\label{eq18}
\mu^{*M} = -bP_{M}(T,\mu),
\end{equation}
\begin{equation}
\label{eq19}
\mu^{*B(\bar B)} = \mu_{B(\bar B)}-bP_{B(\bar B)}(T,\mu)-abn^{2}_{B(\bar B)}+2an_{B(\bar B)},
\end{equation}
where $n_{M}$, $n_{B}$ and $n_{\bar B}$ are the modified number densities of mesons, baryons and anti-baryons, respectively which are given by
\begin{equation}
\label{eq20}
    n_{M}(T,\mu) = \frac{\sum_{i\in M}n_{i}^{id}(T,\mu^{*M})}{1+b\sum_{i\in M}n_{i}^{id}(T,\mu^{*M})},
\end{equation}
\begin{equation}
\label{eq21}
    n_{B(\bar B)}(T,\mu) = \frac{\sum_{i\in B(\bar B)}n_{i}^{id}(T,\mu^{*B(\bar B)})}{1+b\sum_{i\in B(\bar B)}n_{i}^{id}(T,\mu^{*B(\bar B)})}.
\end{equation}

The van der Waals parameters can be obtained by reproducing the ground state of the nuclear matter \cite{Vovchenko:2015vxa} or one can also fit the results for different thermodynamic quantities obtained in lattice QCD and obtain the parameters $``a"$ and $``b"$ \cite{Samanta:2017yhh,Sarkar:2018mbk}. This is interpreted as a VDW-like interaction that mimics the short-ranged residual nuclear force. The parameters in the model are now chosen as $a=0.926$ $GeV fm^{3}$ and $b=(16/3)\pi r^3$, where the hardcore radius $r$ is replaced by $r_{M}=0.2$ $fm$ and $r_{B,(\bar{B})}=0.62$ $fm$, respectively for mesons and (anti)baryons~\cite{Sarkar:2018mbk}. It is also to be noted that although the van der Waals interaction is used in Gibbsian equilibrium, we are using the Boltzmann theory. Such an approach can be justified as both of these coincide for a large number of participants in the system under certain conditions~\cite{mindthegap} {\it i.e.}, provided that the system is in agreement with the Khinchin condition, which describes a phase function having a sufficiently small dispersion of the macro-variable for a system with a large number of constituents. The Boltzmann and Gibbsian equilibrium values agree if,  (i) the macro variable is a sum of the observable in the constituent space, and (ii) the variable in the constituent space corresponds to a partition with cells of equal probability. Assuming we are within these limits, we may apply the Khinchin condition to our case. For example, when we are calculating the total energy, it is actually the sum of the energies of all hadrons, which are equally probable in the given phase space, which is homogeneous in nature. 

Let's now briefly introduce the Boltzmann transport equation in relaxation time approximation in order to estimate
the conductivities and diffusivities of the system under consideration.

\subsection{Electrical and Thermal conductivity from Boltzmann transport equation}
\label{ss2}\  To estimate the transport coefficients of a hadron gas, we take advantage of the Boltzmann Transport Equation (BTE) which is given as \cite{BTE,BTE2}
\begin{equation}
    \label{eq22}
    p^{\mu}\partial_{\mu}f_{i}(x,p) + q_{i}F^{\nu\rho}p_{\rho}\frac{\partial}{\partial p^{\nu}}f_{i}(x,p) = {\cal C}_{i}[f_{i}].
\end{equation}
Here, the $F^{\nu\rho}$ is the electromagnetic field strength tensor and $q_{i}$ is the charge of $i^{th}$ hadronic species. ${\cal C}_{i}[f_{i}]$ is the collision integral that gives the rate of change of the non-equilibrium distribution function, $f_{i}$, due to collisions. At the beginning, the system will be away from equilibrium and with subsequent collisions between the constituent hadrons, it will attain local equilibrium after a certain relaxation time. The collision integral within the relaxation time approximation (RTA) can be given as,
\begin{equation}
    \label{eq23}
    {\cal C}_{i}[f_{i}] \simeq -\frac{p^{\mu}v_{\mu}}{\tau_{i}}(f_{i} - f_{i}^{0}) = -\frac{p^{\mu}v_{\mu}}{\tau_{i}}\delta f_{i}.
\end{equation}
Here, $\tau_{i}$ denotes the system relaxation time, which is of the order of collision time, $v_{\mu}$ is the four velocity and $f_{i}^{0}$ is the equilibrium distribution function, which is defined as;
\begin{equation}
    \label{eq24}
    f_{i}^{0} = \frac{1}{\exp(\frac{E_{i}-\mu}{T})\pm 1},
\end{equation}

%

where $E_{i} = \sqrt{p^{2} + m^{2}}$ is the free energy of the $i^{th}$ hadronic species. When the small departure of the system from equilibrium is assumed to be space-time translational invariant, the first term in Eq.~\ref{eq22} simply vanishes. Assuming only a time directional four-velocity, {\it i.e.}, in the local rest frame, the fluid four-velocity is of the form $v_{\mu} = (1, {\bf{0}})$, so that $p^{\mu}v_{\mu} = p^{0}$. Then considering a constant electric field, the Eq.~\ref{eq22} can be written as
\begin{equation}
    \label{eq25}
    q_{i}\bigg (p_{0}{\bf E} . \frac{\partial f^{0}_{i}}{\partial \bf p} + {\bf E.}{\bf p }\frac{\partial f^{0}_{i}}{\partial p^{0}} \bigg ) = -\frac{p^{0}}{\tau_{i}}\delta f_{i}.
\end{equation}
In the Boltzmann approximation, this can be solved to give
\begin{equation}
    \label{eq26}
    \delta f_{i} = \frac{q_{i}\tau_{i}}{T} \frac{\bf E.\bf p}{p^{0}}f_{i}^{0}.
\end{equation}

Electrical conductivity ($\sigma_{el}$) quantifies the response of a system to an applied electric field.
\begin{equation}
    \label{eq27}
    {\bf j_{el}} = \sigma_{el}{\bf E},
\end{equation}
where ${\bf j_{el}}$ is the current density and ${\bf E}$ is the applied electric field. The electric four current $j^{\mu}_{el}$ is defined as
\begin{equation}
    \label{eq28}
    j^{\mu}_{el} = \sum_{i}q_{i}g_{i}\int \frac{d^{3}p}{(2\pi)^{3}E_{i}}p^{\mu}f_{i}(x,p),
\end{equation}
where $q_{i}$ and $f_{i}(x,p)$ are the charge and the distribution function of the $i^{th}$ hadron. For a small perturbation from equilibrium, the four currents can be approximated as
\begin{equation}
    \label{eq29}
    j^{\mu}_{el} = (j^{\mu}_{0})_{el} + \Delta j^{\mu}_{el},
\end{equation}
where
\begin{equation}
    \label{eq30}
    \Delta j^{\mu}_{el} = \sum_{i} q_{i}g_{i}\int \frac{d^{3}p}{(2\pi)^{3}E_{i}}p^{\mu}\delta f_{i}(x,p), 
\end{equation}
Substituting the value of $\delta f_{i}$ from Eq.~\ref{eq26} into the above equation and hence using the definition of electrical conductivity we get \cite{Kadam:2017iaz}
\begin{equation}
    \label{eq31}
    \sigma_{el} = \frac{1}{3T}\sum_{i}g_{i}\tau_{i}q_{i}^{2}\int \frac{d^{3}p}{(2\pi)^{3}}\frac{\bf p^{2}}{E_{i}^{2}}\times f_{i}^{0},
\end{equation}

where $q_{i}$ and $\tau_{i}$ are the electronic charge and average relaxation time of $i^{th}$ hadronic species and $f_{i}^{0}$ is the equilibrium distribution function given by Eq.~\ref{eq24}.

The relaxation time given above is averaged over all particles and is given by,
\begin{equation}
    \label{eq32}
    \Tilde{\tau}^{-1}_i = \sum_{j}n_j\langle\sigma_{ij}v_{ij}\rangle.
\end{equation}
Here, $n_j$ is the number density of $j^{th}$ hadronic species. The thermal average of total cross-section times relative velocity i.e. $\langle\sigma_{ij}v_{ij}\rangle$ can be calculated as ~\cite{Kadam2015,Cannoni2014,Gondolo1991}
\begin{equation}
    \label{eq33}
    \langle\sigma_{ij}v_{ij}\rangle = \frac{\sigma_{ij}\int_{}^{} d^3p_id^3p_jv_{ij}f^0_if^0_j}{\int^{}_{} d^3p_id^3p_jf^0_if^0_j},
\end{equation}
where we have assumed $\sigma_{ij} = \pi (r_{i}+r_{j})^2$ with $r$ being the corresponding mesonic or baryonic (anti-baryonic) hard core radius. For the same hadronic species, it is reduced to $\sigma = 4\pi r^{2}$. The general expression for a scattering between two different particle species [$i(p_i) + j(p_j) \xrightarrow{} i(p_k) + j(p_l)$] can be given as~\cite{Kadam2015}
\begin{widetext}
\begin{equation}
\label{eq34}
    \langle\sigma_{ij}v_{ij}\rangle = \frac{\sigma_{ij}}{8Tm_i^2m_j^2K_2(\frac{m_i}{T})K_2(\frac{m_j}{T})}\int_{(m_i+m_j)^2}^{\infty}ds\frac{s-(m_i-m_j)^2}{\sqrt{s}}[s-(m_i+m_j)^2]K_1\Big(\frac{\sqrt{s}}{T}\Big).
\end{equation}
\end{widetext}
$s=(p_i+p_j)^2$ denotes the usual Mandelstam variable, $m_i$'s the particle masses and $K_n$'s are modified Bessel functions of order $n$. Knowing the thermal averaged cross section, we calculate the relaxation time using Eq.\ref{eq34}.

Another transport coefficient, Thermal conductivity ($\kappa$), expresses the ability of an interacting system to conduct heat. In the absence of any external field the Boltzmann Transport equation can be written as \cite{BTE2,Kadam:2017iaz}
\begin{equation}
    \label{eq35}
    p^{\mu}\partial_{\mu}f_{i}(x,p) = -\frac{p^{\mu}v_{\mu}}{\tau_{i}}\delta f_{i}.
\end{equation}

The energy momentum tensor and the baryon four current are respectively given as
\begin{equation}
    \label{eq36}
   T^{\mu\nu} = \sum_{i} g_{i}\int \frac{d^{3}p}{(2\pi)^{3}}\frac{p^{\mu}p^{\nu}}{E_{i}}f_{i}(x,p), 
\end{equation}
    
\begin{equation}
    \label{eq37}
    j^{\mu}_{B} = \sum_{i}g_{i}t_{i} \int \frac{d^{3}p}{(2\pi)^{3}}\frac{p^{\mu}}{E_{i}}f_{i}(x,p),
\end{equation}

where, the $g_{i}$ denotes the degeneracy and $t_{i }$ is the baryonic charge of $i^{th}$ hadronic species. Again for a small perturbation, when the system is slightly away from the equilibrium distribution function, The energy momentum tensor becomes 
\begin{equation}
    \label{eq38}
   \Delta T^{\mu\nu} = \sum_{i} g_{i}\int \frac{d^{3}p}{(2\pi)^{3}}\frac{p^{\mu}p^{\nu}}{E_{i}}\delta f_{i}(x,p), 
\end{equation}
and under RTA, the $\delta f_{i}$ can be calculated from the collision term of BTE using Eq.~\ref{eq35}. 
Hence the quantity $\Delta T^{\mu\nu}$ becomes
\begin{equation}
    \label{eq39}
   \Delta T^{\mu\nu} = -\sum_{i} g_{i}\int \frac{d^{3}p}{(2\pi)^{3}}\frac{p^{\mu}p^{\nu}}{E_{i}} \frac{\tau_{i}}{p.v}p^{\rho}\partial_{\rho}f_{i}(x,p).
\end{equation}
Similarly, the change in four current, $\Delta j^{\mu}_{B}$ can be written as,
\begin{equation}
    \label{eq40}
    \Delta j^{\mu}_{B} = -\sum_{i}g_{i}t_{i} \int \frac{d^{3}p}{(2\pi)^{3}}\frac{p^{\mu}}{E_{i}}\frac{\tau_{i}}{p.v}p^{\rho}\partial_{\rho}f_{i}(x,p).
\end{equation}
The derivative $\partial_{\mu}$ can be written in terms of its components parallel and orthogonal to $v^{\mu}$ as \cite{BTE2}
\begin{equation}
    \label{eq41}
    \partial_{\mu} = v_{\mu}D + \nabla_{\mu},
\end{equation}
where, local four-velocity is again considered as $v^{\mu} = (1,\bf 0)$.
\begin{equation}
    \label{eq42}
    D = v^{\mu}\partial_{\mu} = (\partial_{t},\bf 0),
\end{equation}
\begin{equation}
    \label{eq43}
    \nabla_{\mu} = \partial_{\mu} - v_{\mu}D = (0,\partial_{i}).
\end{equation}
Then one can use the conservation equations $\partial_{\mu}T^{\mu\nu} = 0, \partial_{\mu}j^{\mu}_{B} = 0$ to have the following form,
\begin{equation}
    \label{eq44}
    (\epsilon + P)Dv^{\mu} - \nabla^{\mu}P = 0,
\end{equation}
\begin{equation}
    \label{eq45}
    Dn_{net} + n_{net}\nabla_{\mu}v^{\mu} = 0.
\end{equation}
Using these two relations above, we can write the $\Delta T^{\mu\nu}$ and $\Delta j^{\mu}_{B}$ as 

\begin{align}
    \label{eq46}
    \begin{split}
    \Delta T^{\mu\nu} & = \sum_{i} g_{i}\int \frac{d^{3}p}{(2\pi)^{3}E_{i}}\frac{p^{\mu}p^{\nu}}{pv}\frac{1}{T}\tau_{i}f^{0}_{i}\\
    & \times\bigg[pv\Big (\frac{\partial P}{\partial\epsilon}\Big )_{n_{net}}\nabla_{\alpha}v^{\alpha} + p^{\alpha}X_{\alpha} + \frac{p^{\alpha}p^{\beta}}{pv}\nabla_{\alpha}v_{\beta}\\ 
    &+ \Big (\frac{\partial P}{\partial n_{net}} \Big )_{\epsilon}\nabla_{\alpha}v^{\alpha} - \frac{\epsilon + P}{n_{net}}\frac{p^{\alpha}}{pv}X_{\alpha}\bigg ],
    \end{split}
\end{align}
and
\begin{align}
    \label{eq47}
    \begin{split}
    \Delta j^{\mu}_{B} & = \sum_{i} g_{i}\int \frac{d^{3}p}{(2\pi)^{3}E_{i}}\frac{p^{\mu}}{pv}\frac{1}{T}\tau_{i}f^{0}_{i}\\
    & \times\bigg[pv\Big (\frac{\partial P}{\partial\epsilon}\Big )_{n_{net}}\nabla_{\alpha}v^{\alpha} + p^{\alpha}X_{\alpha} + \frac{p^{\alpha}p^{\beta}}{pv}\nabla_{\alpha}v_{\beta}\\ 
    & + \Big (\frac{\partial P}{\partial n_{net}} \Big )_{\epsilon}\nabla_{\alpha}v^{\alpha} - \frac{\epsilon + P}{n_{net}}\frac{p^{\alpha}}{pv}X_{\alpha}\bigg ],
    \end{split}
\end{align}
where
\begin{equation}
    \label{eq48}
    X_{\alpha} = \frac{\nabla_{\alpha}P}{\epsilon + P} - \frac{\nabla_{\alpha}T}{T},
\end{equation}
and $\epsilon$ and $n_{net}$ are the energy density and net baryon number density.
Within Eckart condition one can define $v^{i}$ as \cite{BTE2}
\begin{equation}
    j^{i}_{B} = n_{net}v^{i} + \Delta j^{i}_{B} = 0.\nonumber
\end{equation}
so that the energy flux component $T^{0i}$ becomes
\begin{align}
    \label{eq49}
    \begin{split}
    T^{0i} & = (\epsilon + P)v^{i} + \Delta T^{0i} \\ 
    & = \Delta T^{0i} - \frac{\epsilon + P}{n_{net}}\Delta j^{i}_{B}\equiv I^{i},
    \end{split}
\end{align}
the $I^{i}$ being the heat current, the $\Delta T^{0i}$ and the $\Delta j^{i}_{B}$ is given by
\begin{equation}
    \label{eq50}
    \Delta T^{0i} = \sum_{i} g_{i}\int \frac{d^{3}p}{(2\pi)^{3}}\frac{\bf p^{2}}{3T}\tau_{i}f^{0}_{i}\Big (1 - \frac{\epsilon + P}{n_{net}E_{i}}\Big )X_{i},
\end{equation}
and
\begin{equation}
    \label{eq51}
    \Delta j^{i}_{B} = \sum_{i} t_{i}g_{i}\int \frac{d^{3}p}{(2\pi)^{3}}\frac{\bf p^{2}}{3TE_{i}}\tau_{i}f^{0}_{i}\Big ((1 - \frac{\epsilon + P}{n_{net}E_{i}}\Big )X_{i}.
\end{equation}
Using the Eckart condition, thermal conductivity is defined as 
\begin{equation}
    \label{eq52}
    I^{i} = -\kappa \big [\partial_{i}T - T\partial_{i}P/(\epsilon+P)\big ]. 
\end{equation}
One can now obtain the expression for thermal conductivity using Eq.~\ref{eq50} and Eq.~\ref{eq51} as
\begin{equation}
    \label{eq53}
    \kappa = \frac{1}{3T^{2}}\sum_{i}g_{i}\tau_{i}\int \frac{d^{3}p}{(2\pi)^{3}}\frac{\bf p^{2}}{E_{i}^{2}}\Big(E_{i}-\frac{t_{i}\omega}{n_{net}}\Big)^{2} \times f_{i}^{0} ,
\end{equation}
where $\tau_{i}$ is the relaxation time defined in Eq.~\ref{eq33}. The baryonic charge of $i^{th}$ hadronic species is denoted by $t_{i}$ whereas $\omega = \epsilon_i + P_i$ gives the enthalpy of the $i^{th}$ hadronic species and $n_{net}$ is the net baryon density in the system.
\begin{figure*}[ht!]
\begin{center} 
\includegraphics[scale = 0.44]{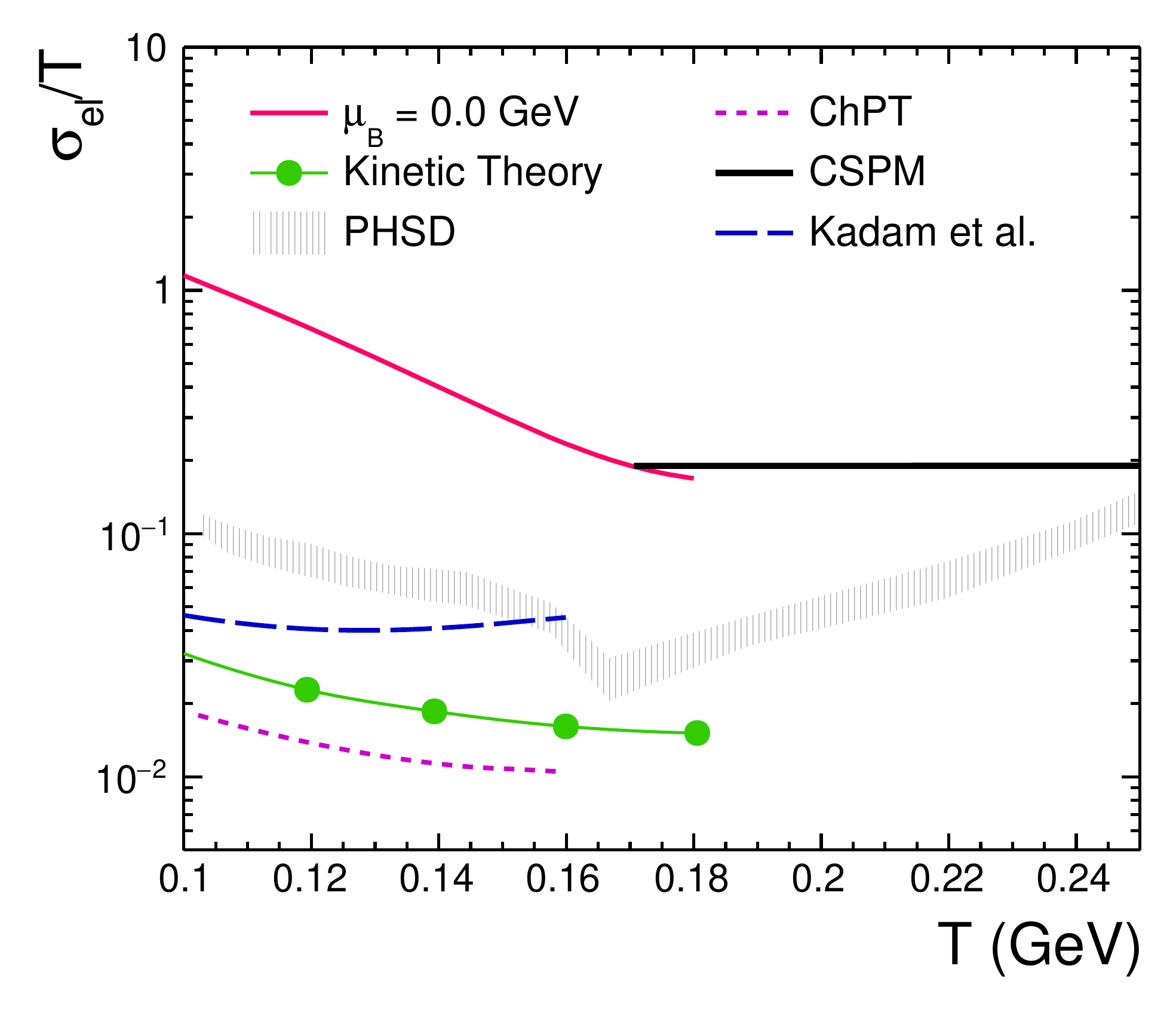}
\includegraphics[scale = 0.44]{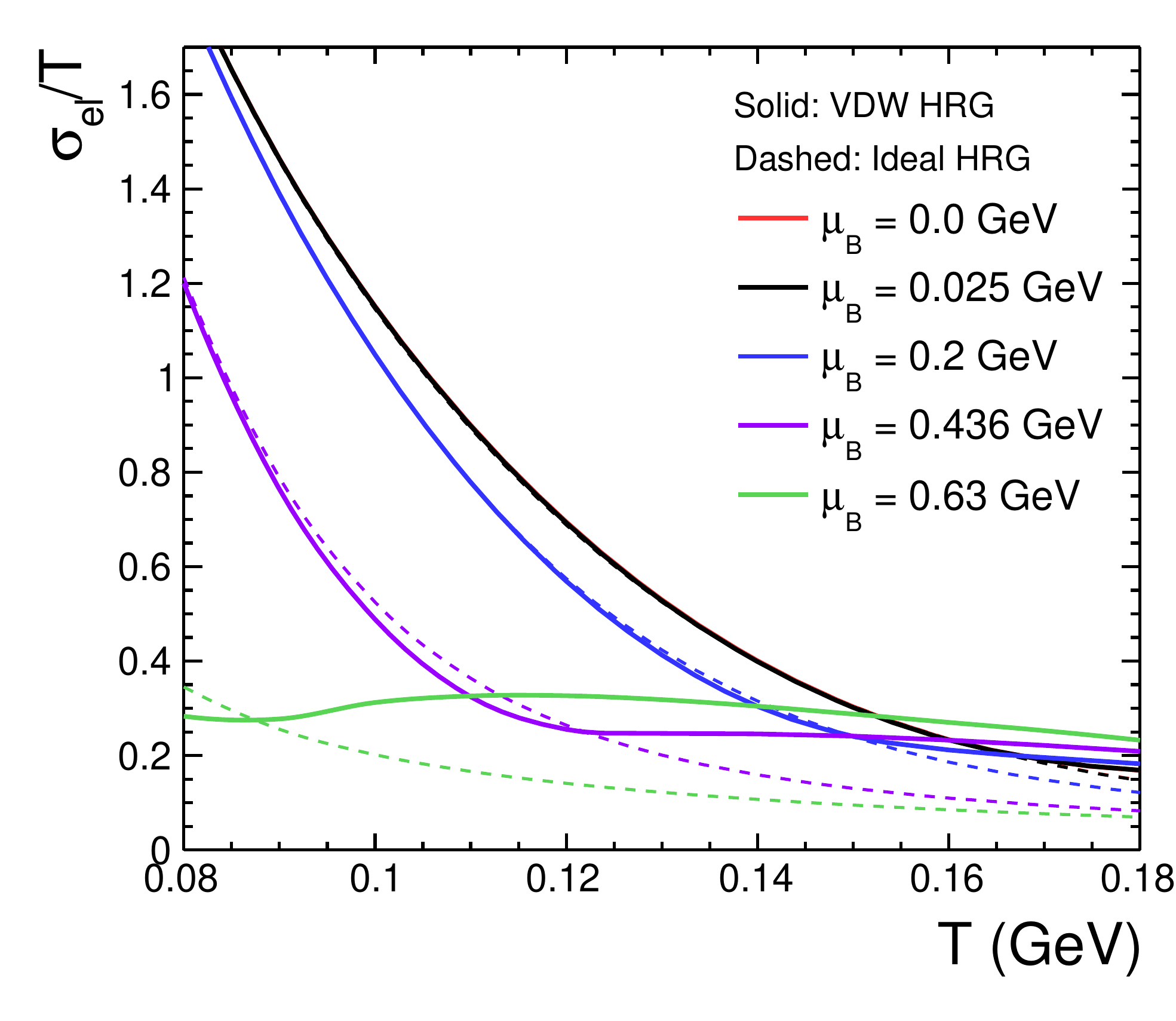}
\caption{(Color Online) (Left) The scaled electrical conductivity as a function of temperature. The solid black line is obtained from CSPM \cite{Sahoo:2018dxn} and the grey band is the result from PHSD \cite{Cassing:2013iz}. The kinetic theory calculations \cite{Kinetic} are represented by green circles. The dotted magenta line is from chiral perturbation theory results \cite{ChPT}, and the dashed blue line shows the result from excluded volume HRG model \cite{Kadam:2017iaz}. (Right) Scaled electrical conductivity is plotted as a function of temperature for different baryon chemical potentials under both HRG and VDWHRG scenarios.}
\label{fig1}
\end{center}
\end{figure*}

The Eq.~\ref{eq31} and Eq.~\ref{eq53} give the expression for the electrical and thermal conductivity respectively, where the quantities like net baryon density, $n_{net}$, enthalpy, $\omega = (\epsilon + P)$, relaxation time, $\tau$, etc. are modified from ideal HRG on account of VDW interaction. The ratio of these two transport properties can now be used to study the Wiedemann-Franz law in a hadronic medium.

\subsection{Diffusivity}
\label{ss4}
In simple words, diffusion is a process of net movement of matter or energy due to the presence of a concentration gradient and it flows from a higher concentration to the lower one. Diffusivity is then defined as the rate of this diffusion. The charge diffusion coefficient which represents the rate of movement of charge in the direction of the concentration gradient is related to the electrical conductivity through the relation \cite{Aarts:2014nba}
\begin{equation}
    \label{eq54}
    D_{el} = \frac{\sigma_{el}}{\chi_{Q}},
\end{equation}
where the $\chi_{Q}$ is the electrical charge susceptibility quantifying the fluctuations of electric charges. Conserved charge susceptibilities can be calculated by the second (or higher) order derivative of the partition function or the equivalent pressure with respect to a chemical potential corresponding to the concerned conserved charge \cite{Samanta:2017yhh}. Electrical charge susceptibility is given by,
\begin{equation}
    \label{eq55}
    \chi_{Q} = \frac{\partial^{2} (P(T,\mu)/T^{4})}{\partial( \mu_{Q}/T)^{2}}.
\end{equation}
\begin{figure*}[ht!]
\begin{center}
\includegraphics[scale = 0.44]{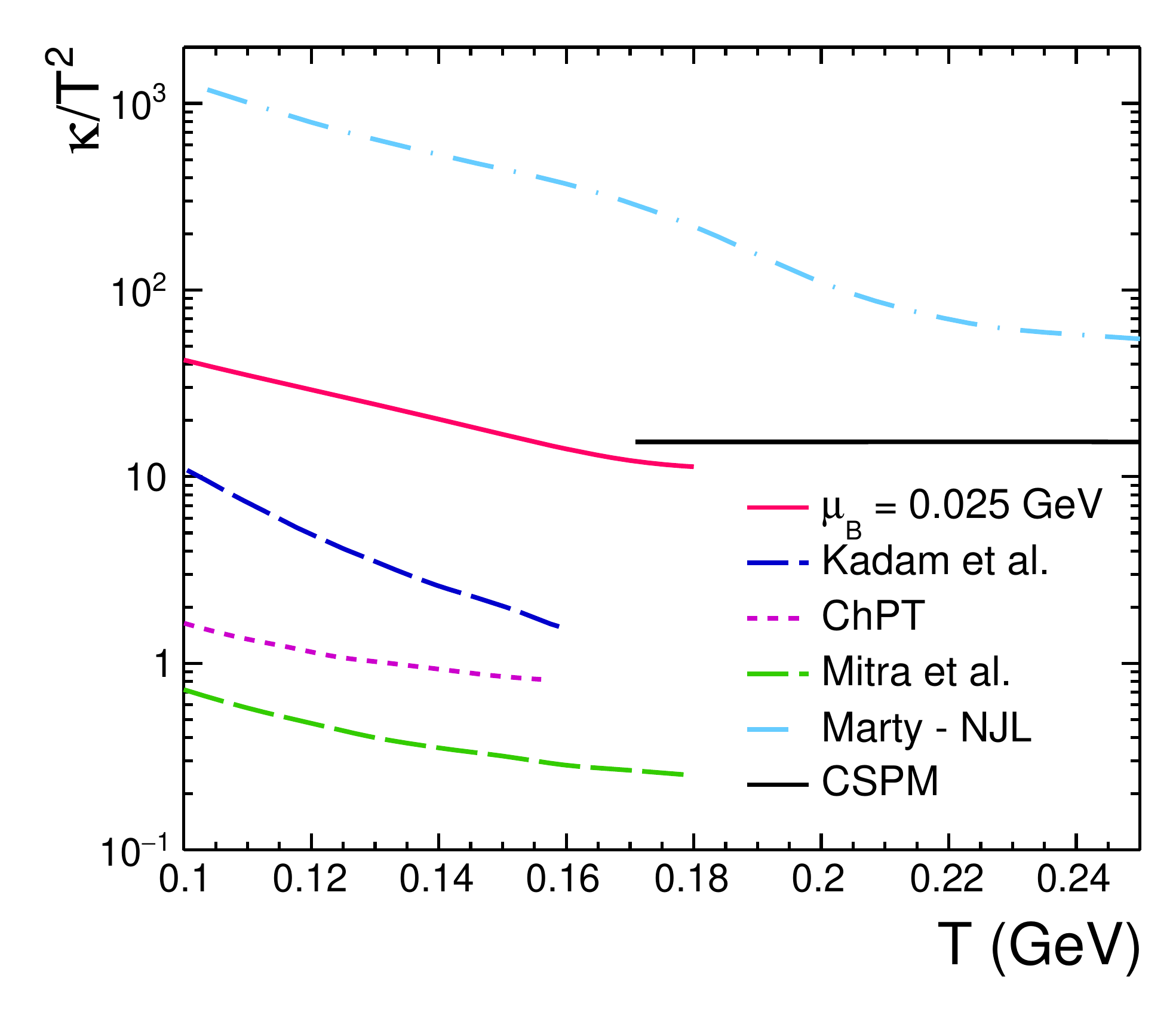}
\includegraphics[scale = 0.44]{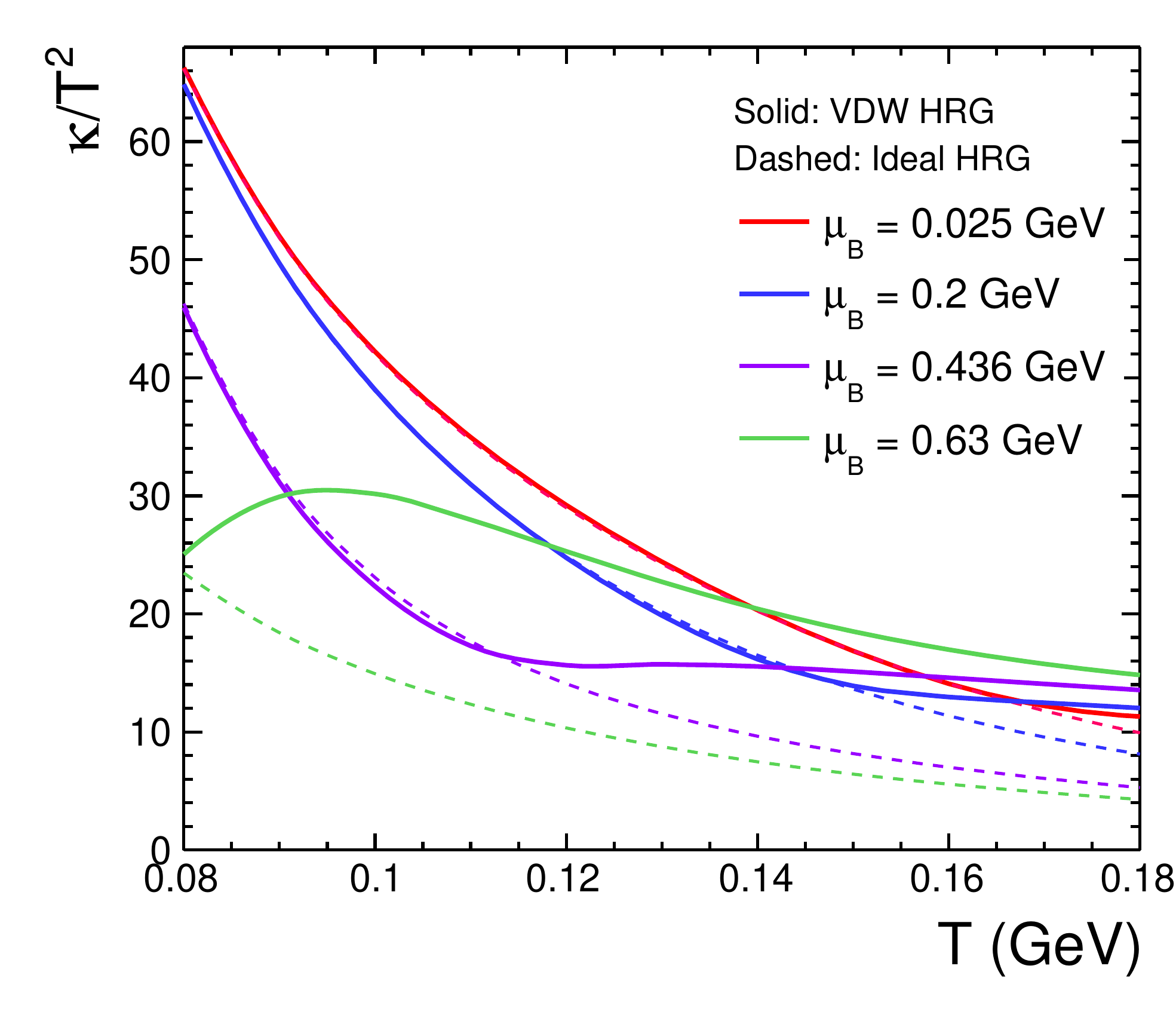}
\caption{(Color Online) Scaled thermal conductivity as a function of temperature. (Left panel) The VDW HRG model at $\mu_{B}=0 $ GeV is compared to results obtained from different models. The black solid line shows the results obtained in CSPM \cite{Sahoo:2018dxn}. The blue dashed line gives the calculations from NJL \cite{Marty:2013ita} model. The dotted magenta line is from chiral perturbation theory \cite{ChPT} and the dashed blue line shows the result from excluded volume HRG model \cite{Kadam:2017iaz}. The green dashed line is for pion gas \cite{mitra}. (Right panel) Scaled thermal conductivity is shown as a function of temperature for different baryon chemical potentials under both HRG and VDWHRG scenarios.}
\label{fig2}
\end{center}
\end{figure*}

Thermal diffusivity is defined as thermal conductivity divided by specific heat capacity and density. Again the specific heat capacity can be written in terms of the derivative of entropy with temperature and hence the thermal diffusivity is given by \cite{heq}
\begin{equation}
    \label{eq56}
    D_{Th} = \kappa\bigg/\Big(\frac{T}{V}\frac{\partial S}{\partial T}\Big).
\end{equation}

\section{Results and Discussion}
\label{results}

We have considered the VDWHRG model, which includes the excluded volume effect of mesons along with both attractive and repulsive interactions between pairs of (anti)baryons. All identified hadrons and resonances up to a mass cut-off of 2.25 GeV \cite{PDG2016} are included in the calculations. The main goal of this work is to estimate the conductivities and corresponding diffusivities (for which we have used the Boltzmann transport equation) and study their dependencies on both temperature and net baryon density.
Five different values of baryon chemical potential are chosen for the analysis; $\mu_{B} = 0.0$ GeV corresponding to LHC energies, $\mu_{B} = $ 0.025, 0.2 GeV corresponding to RHIC at $\sqrt{s_{NN}} = $ 200 GeV $\&$ $\sqrt{s_{NN}} = $ 19.6 GeV, $\mu_{B} = $ 0.436 and 0.630 GeV corresponding to RHIC/FAIR at $\sqrt{s_{NN}} = $ 7.7 GeV and NICA at $\sqrt{s_{NN}} = $ 3 GeV, respectively \cite{Tawfik2016,Munzinger2001,Cleymans2006,Khuntia2019}.

The left panel of Fig.~\ref{fig1} shows the temperature dependence of scaled electrical conductivity, $\sigma_{el}/T$, in the VDWHRG formalism (solid red line). We observe that electrical conductivity decreases with increasing temperature. An increase in temperature increases the number density, decreasing relaxation time due
to more frequent collisions. The randomness in the system is thus larger in the high-temperature scenario. This restricts the charge flow in one particular direction, thus reducing $\sigma_{el}$. Similar behavior is observed for other models in the hadronic phase. We have compared our results with results from parton-hadron-string dynamics (PHSD) transport approach \cite{Cassing:2013iz}, Excluded volume HRG (EVHRG) model \cite{Kadam:2017iaz}, Kinetic theory \cite{Kinetic}, Chiral Perturbation theory (ChPT) \cite{ChPT} and the Color String Percolation Model (CSPM) \cite{Sahoo:2018dxn}. In the PHSD model, $\sigma_{el}$ decreases with temperature in the hadronic phase, attaining a minimum around the critical temperature $T_{c}$, again increasing in the partonic phase. It is noteworthy that there is a smooth transition from our results to the QCD-inspired CSPM results near the critical temperature limit.

On the right panel of Fig.~\ref{fig1} the variation of scaled electrical conductivity is shown as a function of temperature for various $\mu_{B}$ in both ideal and VDWHRG models. $\sigma_{el}/T$ decreases monotonically with temperature in both ideal and VDWHRG cases at low $\mu_{B}$ with VDWHRG results being higher than ideal HRG values at high temperatures. For $\mu_{B} = 0.0 $ GeV, the number density of both mesons and baryons at low temperatures is small, and both models behave similarly. At higher temperatures, the number density increases rapidly and the VDW repulsive part becomes more prominent. This reduces the number of particles per unit volume and in turn, leads to an increase in relaxation time as compared to the ideal HRG scenario. Hence, there is a finite increase in the magnitude of $\sigma_{el}/T$ when a VDW type of interaction is involved. As $\mu_{B}$ gradually increases baryon number density increases and the deviation from the ideal scenario shifts towards lower temperatures. In the regime of low temperature and high $\mu_{B}$, the attractive parameter dominates over the repulsive parameter; thus, the number density becomes higher than the ideal case. As temperature increases, this situation is reversed, resulting in a larger $\sigma_{el}/T$. Once the number density saturates at a higher temperature and particles gain higher energies, this results in a gradual increase in $\sigma_{el}/T$ with an increase in temperature.

The left panel of Fig.\ref{fig2} shows the temperature dependence of scaled thermal conductivity in the VDWHRG model at $\mu_{B} = 0.025$ GeV (solid red line). The $\kappa/T^{2}$ also shows a similar kind of temperature dependency as shown by the electrical conductivity in Fig.~\ref{fig1}. 
On the increasing temperature, the number density in the system increases, which in turn increases the frequency of collisions. These random collisions restrict the flow of heat energy. 
\begin{figure*}[ht!]
\begin{center}
\includegraphics[scale = 0.44]{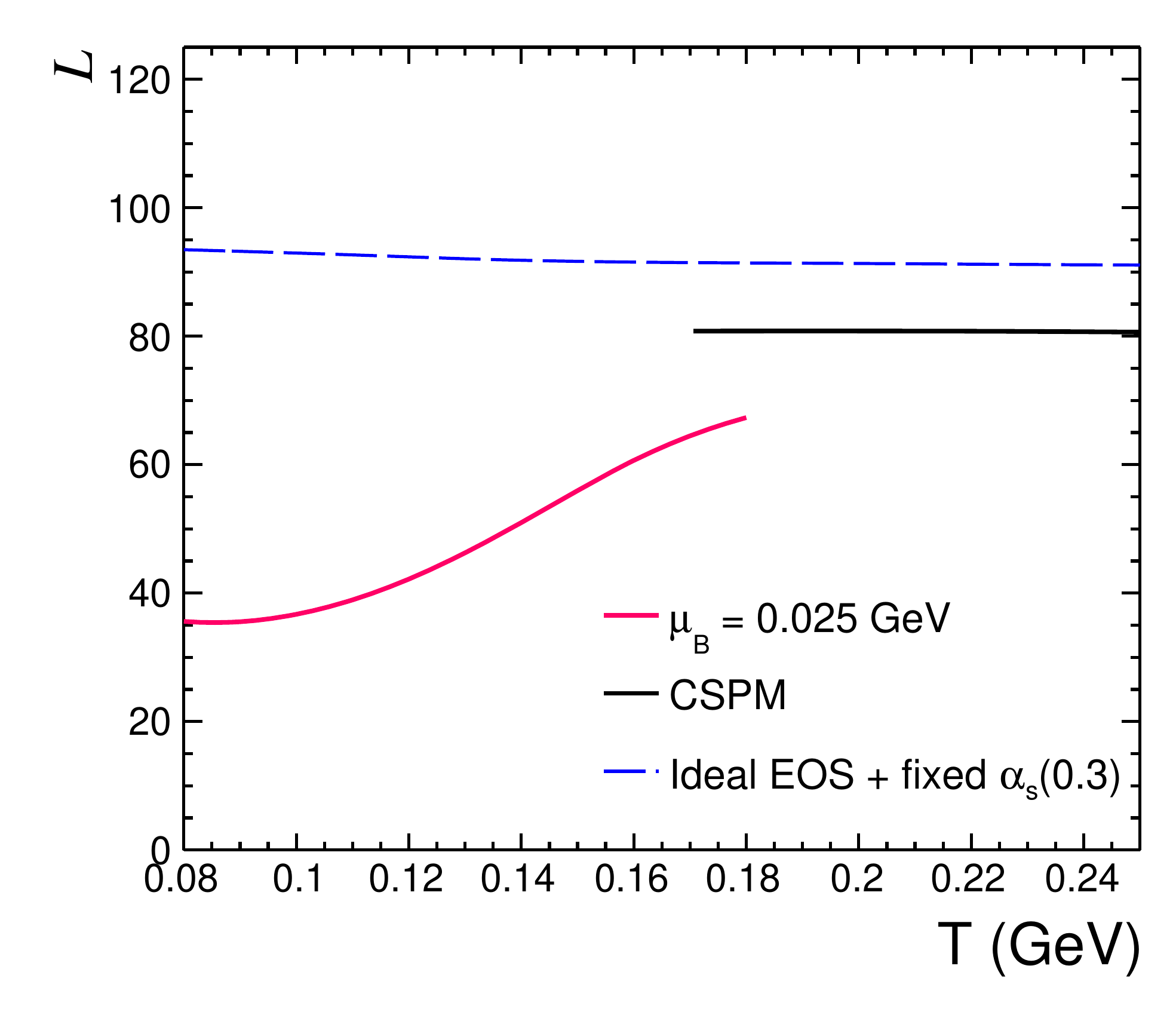}
\includegraphics[scale = 0.44]{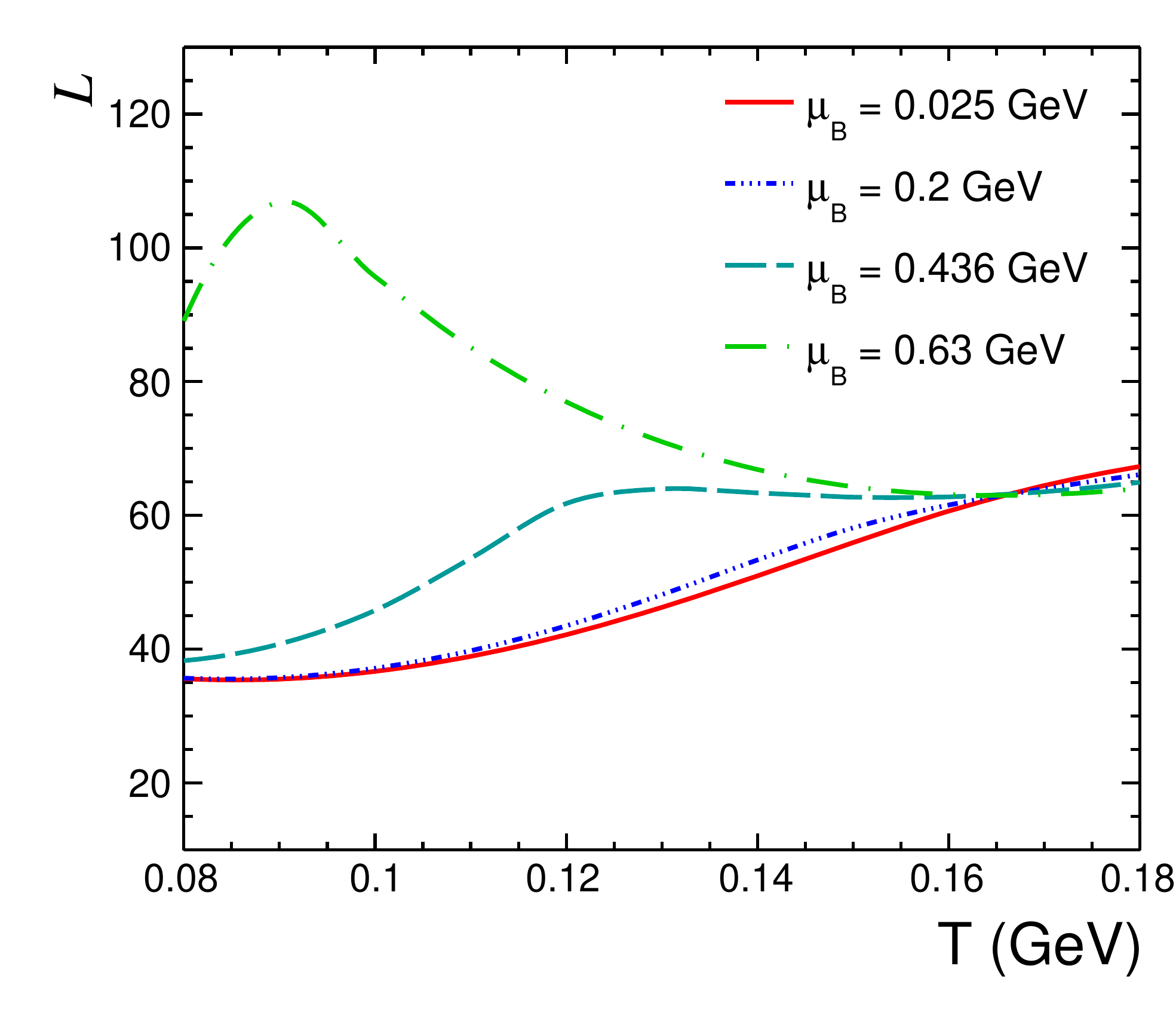}
\caption{(Color Online) (Left panel) The Lorenz number as a function of temperature. The black solid line shows the results from CSPM \cite{Sahoo:2018dxn}. The blue dashed line represents the result of the ideal EOS for a fixed coupling constant \cite{Mitra:2017sjo}. (Right panel) Lorenz number as a function of temperature for different baryon chemical potential.}
\label{fig3}
\end{center}
\end{figure*}
The results obtained are compared with the three quark (u,d,s) Nambu-Jona-Lasino (NJL) model ~\cite{Marty:2013ita}, EVHRG model \cite{Kadam:2017iaz}, ChPT model \cite{ChPT}, CSPM \cite{Sahoo:2018dxn} and pion gas in relaxation time approximation (RTA) \cite{mitra}, all of which have a general decreasing trend of $\kappa/T^{2}$ with an increase in temperature. The difference in magnitude for various models may be because of the fact that in the NJL model, degrees of freedom are calculated at the partonic level, while for the EVHRG model, the difference is mainly due to the particle radii chosen. The low values obtained by Mitra \textit{et. al.} \cite{mitra} for pion gas signifies the importance of contribution from other hadrons. Once again, we note that the QCD-inspired CSPM model agrees well with our results near the possible critical temperature.

Fig.~\ref{fig2} (right panel) also shows the $\mu_{B}$ dependence of $\kappa/T^{2}$. The relaxation time, $\tau$ decreases with increasing number density and thus hinders the transport of thermal energy through the medium. As stated in the case of electrical conductivity, the VDWHRG thermal conductivity deviates from ideal HRG towards higher temperature as a result of lower number density due to repulsive interactions. The overall thermal conductivity however decreases with an increase in temperature. At a particular $\mu_{B}$, because of the van der Waals interaction, the increase in both enthalpy and net baryon density with temperature is reduced as compared to that in the ideal case. However decrease in net baryon density (as compared to the ideal case) is much more pronounced and therefore at high temperatures, thermal conductivity increases resulting in the minima observed. As the baryon chemical potential increases further, this behavior is seen even at lower temperatures and the minima get shifted accordingly.

The response of an electrically charged interacting medium to an external electric field is measured by electrical conductivity. On the other hand, thermal conductivity gives a measure of the transport of thermal energy on account of a temperature gradient. According to the Wiedemann-Franz law, the ratio of thermal to electrical conductivity is proportional to temperature with the proportionality constant being known as the Lorenz number. This law is observed to be valid for metals where the Lorenz number depends on the nature of the metal. On the left panel of fig.~\ref{fig3}, our results at baryon chemical potential, $\mu_{B} = 0.025$ GeV is compared with the results from CSPM \cite{Sahoo:2018dxn} and with the calculations based on an ideal EOS framework with a fixed coupling constant ($\alpha_{s} = $ 0.3) \cite{Mitra:2017sjo}. Both calculations do not show any dependence of \textit{L} on temperature. Our analysis shows that at the low-temperature region, the Lorenz number is nearly constant and it increases almost linearly after a temperature of $T \sim$ 110 MeV, indicating that the Wiedemann-Franz law is violated in a hadron gas at higher temperatures. This also indicates that both electrical and thermal conductivities are proportional at low temperatures. However, with an increase in temperature, thermal conductivity appears to be more prominent in comparison to electrical conductivity. When the results of VDWHRG with hadronic degrees of freedom are combined with that of CSPM results based on the assumption of color degrees of freedom, one observes a change in the behavior of the Lorentz number around $T \sim 170$ MeV. This may hint at a quark-hadron phase transition around $T \sim 170$ MeV. This also goes in line with the lQCD predicted value of a critical temperature for the deconfinement transition \cite{Borsanyi:2010bp}.

\begin{figure*}[ht!]
\includegraphics[scale = 0.44]{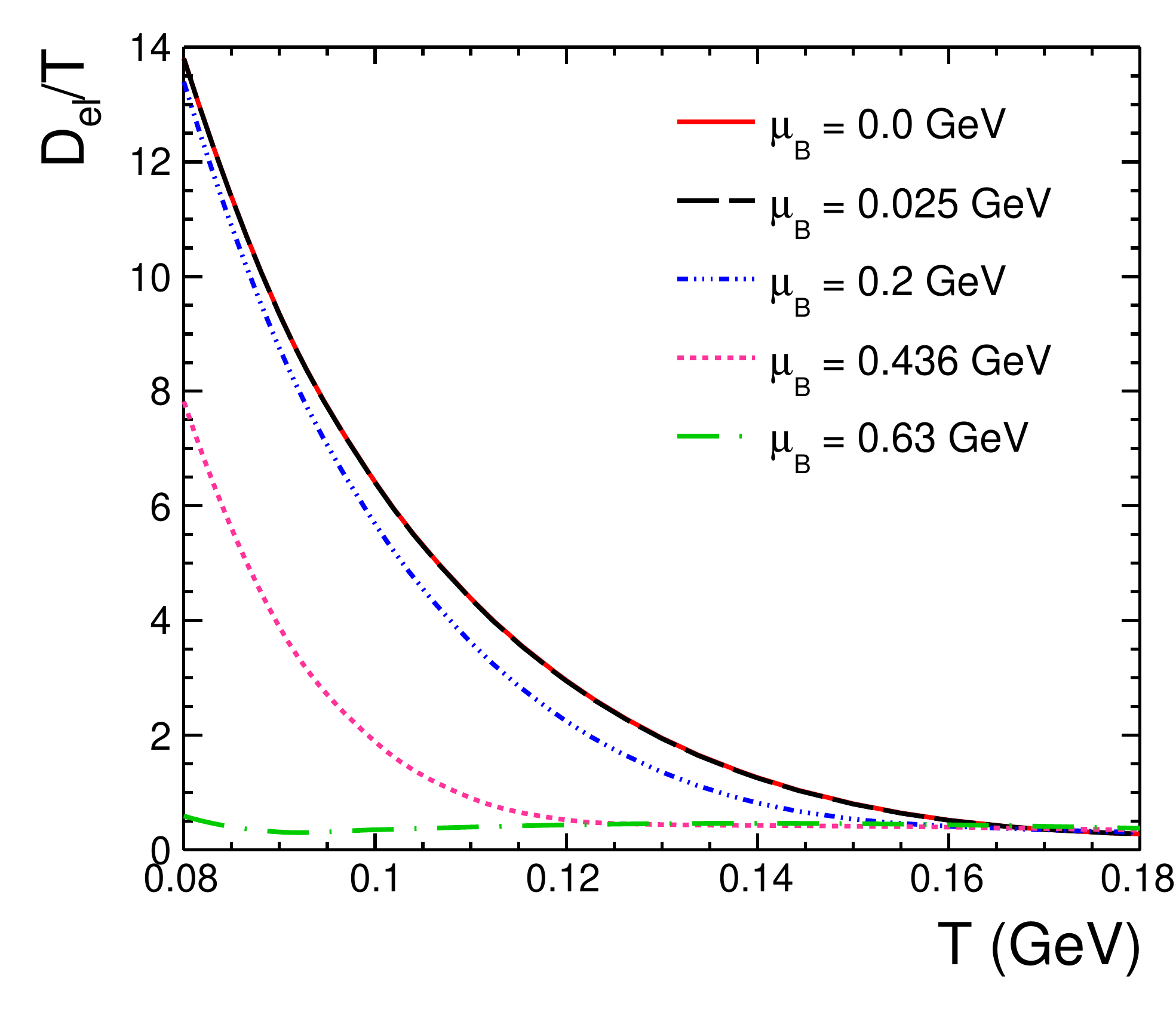}
\includegraphics[scale = 0.44]{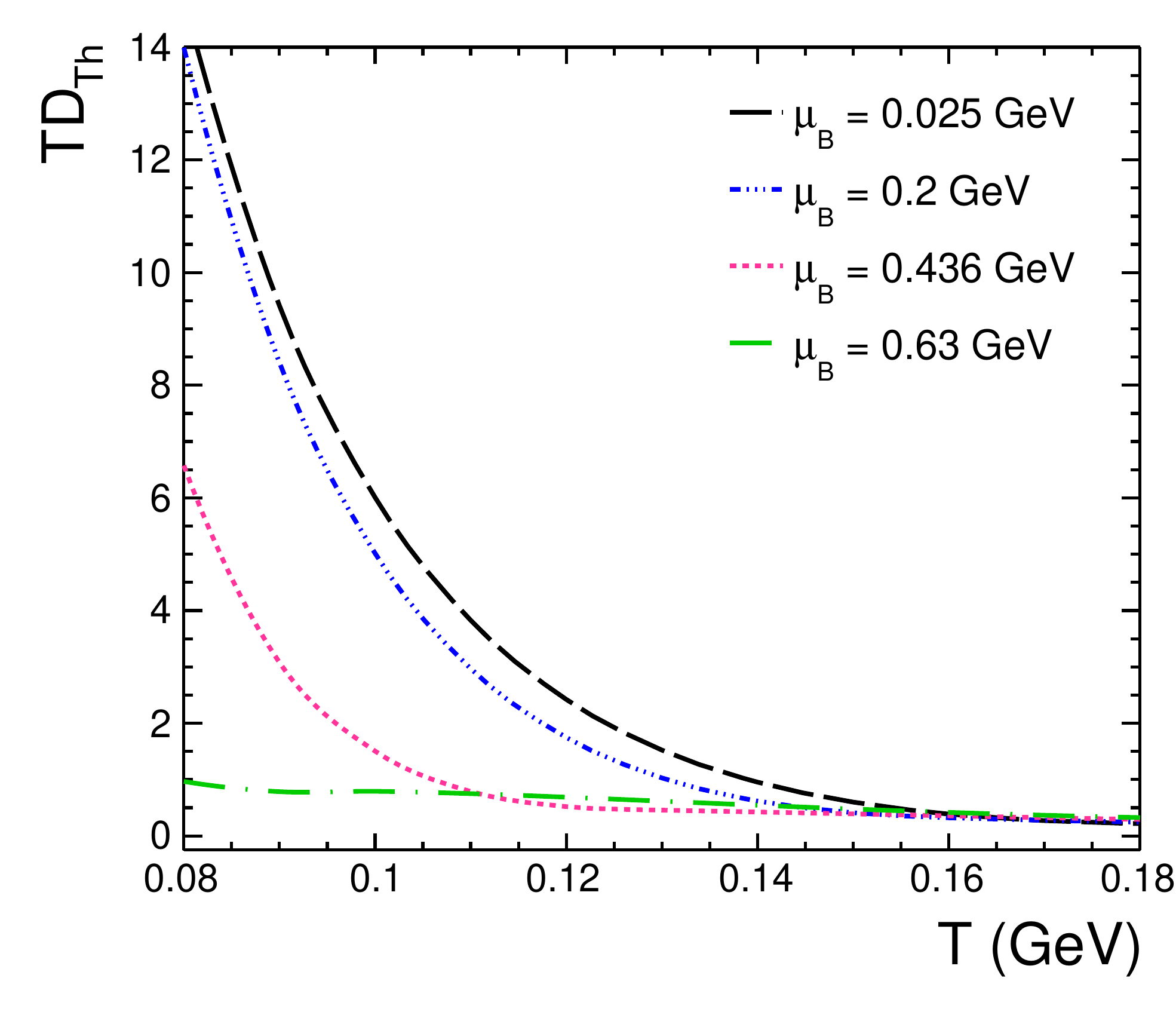}
\caption{(Color Online) The left and right panels show the dimensionless electrical diffusivity and dimensionless thermal diffusivity as functions of temperature for certain baryochemical potentials, respectively.}
\label{fig4}
\end{figure*}

 The right panel of Fig.~\ref{fig3} shows the $\mu_{B}$ dependence of the Lorenz number. As $\mu_{B}$ increases, \textit{L} is seen to increase. 
With the increase in baryon chemical potential, there is a decrease in both electrical and thermal conductivity. However, the decrease in electrical conductivity is more prominent than that of thermal conductivity. Therefore, with increasing $\mu_{B}$, the Lorenz number also increases. Again, at high $\mu_{B}$ and high temperature, there is an enhancement in the thermal conductivity, which reflects in the results of Lorenz number in the same Fig.~\ref{fig3}. At high $\mu_{B}$, the Lorenz number decreases at high temperature after attaining a maximum at low-temperature region. Overall, the Wiedemann-Franz law is violated for all cases of $\mu_{B}$. We also observe the formation of a maximum in the temperature dependence of \textit{L} which shifts towards lower temperature at higher $\mu_{B}$. This behavior may be the signature of a phase transition which might be seen at even lower temperatures if $\mu_{B}$ is further increased. It is, however, to be noted that if such a phase transition occurs, then it would be a liquid-gas phase transition as we have chosen a model based on the van der Waals interactions.

Fig.~\ref{fig4} shows the temperature dependence on thermal and charge diffusivities at various $\mu_{B}$. On the left panel, the dimensionless quantity $D_{el}/T$ is plotted as a function of temperature for different values of $\mu_{B}$. It is observed that $D_{el}/T$ decreases with increasing temperature. When we talk of diffusivity, we essentially study the rate of diffusion of a quantity in the direction of a concentration gradient. One can naively state that a low diffusivity at high temperatures essentially means the system is denser as compared to the low-temperature regime. This behavior is similar to the trend observed in the case of electrical conductivity. Moreover, results from lQCD show minima for electrical diffusivity near the critical temperature, and the trend again increases with an increase in temperature \cite{Aarts:2014nba}. This minimum appears because of the phase transition from the hadronic to the QGP phase.

The same behavior is seen for thermal diffusivity plotted as a function of temperature in the right panel of Fig.~\ref{fig4}. The dimensionless quantity $TD_{Th}$ decreases with temperature. At high temperatures, the increase in number density and collision makes the medium less capable of transferring the thermal energy, and hence the diffusion process slows down at high temperatures. Thus more heat is absorbed than it is transported. Both the diffusivities can be helpful in locating the phase transition because of their minima at the critical temperature.

\section{Summary}
\label{sum}

In this work, we have estimated the electrical and thermal conductivities along with their respective diffusivities. We have studied these quantities as functions of temperature and baryochemical potential. The effect of van der Waals interaction is also studied on the electrical and thermal conductivities. The VDW interaction seems to have a dominant impact on the transport properties at high baryon densities. The diffusivities show universal decreasing trends for lower baryochemical potential. Along with this, we have also studied the Weidemann-Franz law in the VDWHRG model. The hadron gas under VDW interactions seems to be violating the Weidemann-Franz law at a higher temperature regime. It should be noted that for a partonic system \cite{Sahoo:2018dxn,Mitra:2017sjo} the Weidemann-Franz law is respected 
like a system of electrons in a metal. Whereas, when one considers a hadronic system with or without interaction
taking a thermalized Boltzmann distribution or assuming the system is a little away from equilibrium \cite{Rath:2019nne} (one uses Tsallis non-extensive statistics), the Weidemann-Franz law is seen to be violated.

\section*{Acknowledgement}

KP and Ronald Scaria acknowledge the doctoral fellowships from UGC, and CSIR, Government of India, respectively. The authors gratefully acknowledge the DAE-DST, Govt. of India funding under the mega-science project -- “Indian participation in the ALICE experiment at CERN" bearing Project No. SR/MF/PS-02/2021-IITI (E-37123).

 \end{document}